%% file: xxT_arxiv.tex
\documentclass[twoside,11pt]{article}
\usepackage{jmlr2e}
\usepackage{wrapft}
\usepackage[utf8]{inputenc} 
\usepackage[T1]{fontenc}    
\usepackage{hyperref}       
\usepackage{url}            
\usepackage{booktabs}       
\usepackage{amsfonts}       
\usepackage{nicefrac}       
\usepackage{microtype}      
\usepackage{algorithm, algpseudocode}
\usepackage{enumitem}
\setlist[itemize]{leftmargin=2}
\input{./settings_custom.tex}

\usepackage{mathtools}
\usepackage{amsfonts}
\usepackage{amssymb}
\title{Mutual information for symmetric rank-one matrix estimation: A proof of the replica formula}
\author{\name Jean Barbier, Mohamad Dia and Nicolas Macris \email firstname.lastname@epfl.ch \\
\addr Laboratoire de Th\'eorie des Communications, Facult\'e Informatique et Communications,\\
Ecole Polytechnique F\'ed\'erale de Lausanne, 1015, Suisse.
\AND
\name Florent Krzakala \email florent.krzakala@ens.fr \\
\addr Laboratoire de Physique Statistique, CNRS, PSL Universit\'es et
Ecole Normale Sup\'erieure, \\Sorbonne Universit\'es et Universit\'e Pierre \& Marie Curie, 75005, Paris, France.
\AND
\name Thibault Lesieur and Lenka Zdeborov\'a \email lesieur.thibault,lenka.zdeborova@gmail.com \\
\addr Institut de Physique Th\'eorique, CNRS, CEA, Universit\'e Paris-Saclay, \\ F-91191, Gif-sur-Yvette, France.
}
\begin{document}
%
\maketitle
\begin{abstract}
  Factorizing low-rank matrices has many applications in machine
  learning and statistics. For probabilistic models in the Bayes
  optimal setting, a general expression for the mutual information has
  been proposed using heuristic statistical physics computations, and
  proven in few specific cases.  Here, we show how to rigorously prove
  the conjectured formula for the symmetric rank-one case. This allows
  to express the minimal mean-square-error and to characterize the
  detectability phase transitions in a large set of estimation
  problems ranging from community detection to sparse PCA. We also
  show that for a large set of parameters, an iterative algorithm
  called approximate message-passing is Bayes optimal. There exists,
  however, a gap between what currently known polynomial algorithms
  can do and what is expected information theoretically. Additionally,
  the proof technique has an interest of its own and exploits three
  essential ingredients: the interpolation method introduced in
  statistical physics by Guerra, the analysis of the approximate
  message-passing algorithm and the theory of spatial coupling and threshold saturation in
  coding. Our approach is generic and applicable to other open
  problems in statistical estimation where heuristic statistical
  physics predictions are available.
\end{abstract}
Consider the following probabilistic rank-one matrix estimation
problem: one has access to noisy observations
$\tbf{w}\!=\!(w_{ij})_{i,j=1}^n$ of the pair-wise product of the
components of a vector
$\tbf{s}\!=\!(s_1,\dots, s_n)^\intercal \!\in \! \mathbb{R}^n$ with
i.i.d components distributed as $S_i\!\sim\! P_0$, $i\!=\!1,\dots,
n$.
The entries of $\tbf{w}$ are observed through a noisy element-wise
(possibly non-linear) output probabilistic channel
$P_{\rm out}(w_{ij}|s_is_j/\sqrt{n})$. The goal is to estimate the
vector $\tbf{s}$ from $\tbf{w}$ assuming that both $P_0$ and
$P_{\rm out}$ are known and independent of $n$ (noise is symmetric so
that $w_{ij}\!=\!w_{ji}$). Many important problems in statistics and
machine learning can be expressed in this way, such as sparse PCA
[\cite{zou2006sparse}], the Wigner spike model
[\cite{johnstone2012consistency,6875223}], community detection
[\cite{deshpande2015asymptotic}] or matrix completion
[\cite{candes2009exact}].

Proving a result initially derived by a heuristic method from
statistical physics, we give an explicit expression for the mutual
information and the
information theoretic minimal mean-square-error (MMSE) in the asymptotic $n\! \to\! +\infty$ limit. Our results
imply that for a large region of parameters, the 
posterior marginal expectations of the underlying signal components
(often assumed intractable to compute) can be obtained in the leading
order in $n$ using a
polynomial-time algorithm called approximate message-passing
(AMP) [\cite{rangan2012iterative,6875223,deshpande2015asymptotic,lesieur2015phase}]. We also demonstrate the existence of a region where both AMP and spectral methods [\cite{baik2005phase}] fail to provide a good answer to the estimation problem, while it is nevertheless information theoretically possible to do so. We illustrate our theorems with examples and also briefly discuss the implications in terms of computational complexity.
\section{Setting and main results}
\subsection{The additive white Gaussian noise setting}
\label{sec:universal}
A standard and natural setting is the case of additive
white Gaussian noise (AWGN) of known variance $\Delta$,
\begin{align} \label{eq:mainProblem}
w_{ij} = \frac{s_i s_j}{\sqrt{n}} + z_{ij}\sqrt{\Delta} ,
\end{align}
where $\tbf{z}\!=\!(z_{ij})_{i,j=1}^n$ is a symmetric matrix with i.i.d entries $Z_{ij}\! \sim\! \mathcal{N}(0,1)$, $1\!\leq\! i\!\leq\! j\!\leq\! n$. Perhaps surprisingly,
it turns out that this Gaussian setting is sufficient to completely
characterize all the problems discussed in the introduction, even if
these have more complicated output channels. This is made possible by a theorem of channel universality [\cite{krzakala2016mutual}] (already proven for community detection in
[\cite{deshpande2015asymptotic}] and conjectured in [\cite{lesieur2015mmse}]). This theorem states that given an
output channel $P_{\rm out}(w|y)$, such that $\log P_{\rm out}(w|y\!=\!0)$ is three times differentiable with bounded second and
third derivatives, then the mutual information satisfies
$I(\tbf{S};\tbf{W} ) \!= \!I (\tbf{S}; \tbf{S}\tbf{S}^{\intercal}/\sqrt{n}\! +\!\tbf{Z} \sqrt{\Delta}) \!+\!\mathcal{O}(\sqrt{n} )$,
where $\Delta$ is the
inverse Fisher information (evaluated at $y\!=\!0$) of the output channel: $\Delta^{-1}\!\defeq\!\mathbb{E}_{P_{\rm out}(w|0)}[(\partial_y \log P_{\rm out}(W|y)\vert_{y=0})^2]$. Informally, this means that we only have to compute the mutual information for an AWGN channel to take care of a wide range of problems, which can be expressed in terms of their Fisher information. In this paper we derive rigorously, for a large class of signal distributions $P_0$, an explicit one-letter formula for the mutual information per variable $I(\tbf{S} ; \tbf{W})/n$ in the asymptotic limit $n\!\to\!+\infty$.
\subsection{Main result} \label{subsec:mainresult}
Our central result is a proof of the expression for the asymptotic $n\!\to\!+\infty$ mutual
information per variable via the so-called {\em replica symmetric
  potential function} $i_{\rm RS}(E; \Delta)$ defined as
%
\begin{align}\label{eq:potentialfunction}
i_{\rm RS}(E;\Delta) \defeq \frac{(v-E)^2+v^2}{4\Delta} - \mathbb{E}_{S, Z}\biggl[\ln\biggl(\int dx\,P_0(x)
e^{-\frac{x^2}{2\Sigma(E;\Delta)^{2}} 
+ x\bigl(\frac{S}{\Sigma(E;\Delta)^{2}} + \frac{Z}{\Sigma(E;\Delta)}\bigr)}\biggr)\biggr] \, ,
\end{align}
with $Z\!\sim\! \mathcal{N}(0,1)$, $S\!\sim\! P_0$, $\mathbb{E}[S^2]\! =\! v$ and 
$\Sigma(E;\Delta)^2\!\defeq\!\Delta/ (v \! -\!  E)$, $E\!\in\! [0, v]$. 
Here we will assume that $P_0$ is a discrete distribution over 
a finite bounded real alphabet $P_0(s) \!=\! \sum_{\alpha=1}^\nu p_\alpha \delta(s\! - \! a_\alpha)$. 
Thus the only continuous integral in \eqref{eq:potentialfunction} is the Gaussian over $z$. Our results 
can be extended to mixtures of discrete and continuous 
signal distributions at the expense of technical complications in some proofs. 

It turns out that both the information theoretical and algorithmic AMP thresholds are determined by the set of stationary points of \eqref{eq:potentialfunction}
(w.r.t $E$).
It is possible to show that for all $\Delta\!>\!0$ there always exist at least one stationary minimum.
Note $E\!=\!0$ is never a stationary 
point (except for $P_0$ a single Dirac mass) and $E\!=\!v$ is stationary only if $\mathbb{E}[S]\! =\!0$. 
In this contribution 
we suppose that at most three stationary points exist, corresponding to situations with at most one phase transition. 
We believe that situations 
with multiple transitions can also be covered by our techniques.
\begin{theorem}[One letter formula for the mutual information]\label{thm1}
Fix $\Delta\!>\!0$ and assume $P_0$ is a discrete distribution such that $i_{\rm RS}(E;\Delta)$ given by \eqref{eq:potentialfunction} has at 
most three stationary points. 
Then 
 \begin{align}\label{eq:replicaformula}
 \lim_{n\to +\infty}\frac{1}{n} I(\tbf{S} ; \tbf{W})
 = \min_{E\in [0, v]}i_{\rm RS}(E;\Delta).
 \end{align}
\end{theorem}
The proof of the {\it existence of the limit} does not require the above hypothesis on $P_0$. Also, it was first shown
in [\cite{krzakala2016mutual}] that for all $n$,
$I(\tbf{S}; \tbf{W})/n \!\leq\! \min_{E\in [0,
  v]}i_{\rm RS}(E;\Delta)$, an inequality that {\it we will use} in the proof section. 
It is conceptually useful to define the following threshold:
\begin{definition}[Information theoretic threshold]\label{def-delta-opt} 
Define $\Delta_{\rm Opt}$ as the first non-analyticity point of the asymptotic mutual information per variable as $\Delta$ 
increases, that is formally \\ $\Delta_{\rm Opt} \!\defeq\!\sup\{\Delta\vert\lim_{n\to +\infty}I(\tbf{S}; \tbf{W})/n \ \text{is analytic in} \ ]0, \Delta[\}$.
\end{definition}
When $P_0$ is such that \eqref{eq:potentialfunction} has at most three stationary points, as 
discussed below, then $\min_{E\in [0, v]} i_{\rm RS}(E;\Delta)$ has at most one non-analyticity point 
denoted $\Delta_{\rm RS}$ (if $\min_{E\in [0, v]} i_{\rm RS}(E;\Delta)$ is analytic over all $\mathbb{R}_+$ we set $\Delta_{\rm RS} \!=\!+\infty$). 
Theorem~\ref{thm1} gives us a mean to {\it compute} the information theoretical threshold $\Delta_{\rm Opt} \!=\! \Delta_{\rm RS}$.
A basic application of theorem~\ref{thm1} is the expression of the
MMSE:
\begin{corollary}[Exact formula for the MMSE]\label{cor:MMSE}
For all $\Delta\neq \Delta_{\rm RS}$, the 
matrix-MMSE ${\rm Mmmse}_n \!\defeq \!\mathbb{E}_{\tbf{S}, \tbf{W}}\|\tbf{S}\tbf{S}^{\intercal} \!
-\! \mathbb{E}[\tbf{X}\tbf{X}^{\intercal}\vert \tbf{W}]\|_{\rm F}^2/n^2$ ($\|\!-\!\|_{\rm F}$ being the Frobenius norm) 
is asymptotically $\lim_{n\to+\infty} {\rm Mmmse}_n(\Delta^{-1}) \!=\! v^2\!-\!(v\!-\!\argmin_{E\in[0,v]} i_{\rm RS}(E;\Delta))^2$. 
Moreover, if $\Delta \!<\!\Delta_{\rm AMP}$ (where $\Delta_{\rm AMP}$ is the algorithmic threshold, see definition~\ref{algo-thresh-def}) or $\Delta\! >\! \Delta_{\rm RS}$, then 
the usual vector-MMSE ${\rm Vmmse}_n \!\defeq\! \mathbb{E}_{\tbf{S}, \tbf{W}}\|\tbf{S}\! -\! \mathbb{E}[\tbf{X}\vert\tbf{W}]\|_2^2/n$ 
satisfies $\lim_{n\to +\infty}{\rm Vmmse}_n\!=\!\argmin_{E\in[0,v]} i_{\rm RS}(E;\Delta)$.
\end{corollary}
It is natural to conjecture that the vector-MMSE is given by $\argmin_{E\in[0,v]} i_{\rm RS}(E;\Delta)$ for all $\Delta\!\neq \!\Delta_{\rm RS}$, 
but our proof does not quite yield the full statement. 

A fundamental consequence concerns the performance of the AMP algorithm
[\cite{rangan2012iterative}] for estimating $\tbf{s}$.
AMP has been analysed rigorously in
[\cite{BayatiMontanari10,Montanari-Javanmard,deshpande2015asymptotic}]
where it is shown that its asymptotic performance is tracked by {\it state evolution}. Let $E^t\!\defeq \!\lim_{n\to+\infty}\mathbb{E}_{{\tbf S}, {\tbf Z}}[\| {\tbf S} \! - \! \hat{\bs}^{t}\|^2_2]/n$ be the asymptotic average vector-MSE of the AMP estimate $\hat{\bs}^t$ at time $t$. Define ${\rm mmse}(\Sigma^{-2}) \!\defeq\! \mathbb{E}_{S, Z}[( S \!-\! \mathbb{E}[X\vert S \!+\! \Sigma Z])^2]$ as the usual scalar mmse function associated to a scalar AWGN channel of noise variance $\Sigma^2$, with $S\!\sim\! P_0$ and $Z\!\sim\! \mathcal{N}(0,1)$. Then
\begin{align}\label{recursion-uncoupled-SE}
E^{t+1} = {\rm mmse}(\Sigma(E^{t}; \Delta)^{-2}), \qquad E^0 = v,
\end{align}
is the state evolution recursion. Monotonicity properties of the mmse function imply that $E^t$ is a decreasing sequence such that $\lim_{t\to +\infty}E^t \!=\! E^\infty$ exists. Note that 
when $\mathbb{E}[S]=0$ and $v$ is an unstable fixed point, as such, state evolution ``does not start''. While this is not really a problem when one runs AMP in practice, for analysis purposes one can slightly bias $P_0$ and remove the bias at the end of the proofs.
%
%
\begin{definition}[AMP algorithmic threshold]\label{algo-thresh-def}
For $\Delta\!>\!0$ small enough, the fixed point equation corresponding to 
\eqref{recursion-uncoupled-SE} has a unique solution for all noise values in $]0, \Delta[$. We define 
$\Delta_{\rm AMP}$ as the supremum of all such $\Delta$. 
\end{definition}
\begin{corollary}[Performance of AMP]\label{perfamp}
  In the limit $n\!\to\!+\infty$, AMP initialized without any knowledge other than $P_0$ yields upon convergence the asymptotic matrix-MMSE as well as the asymptotic vector-MMSE iff
  $\Delta \!<\!\Delta_{\rm AMP}$ or $\Delta\! >\! \Delta_{\rm RS}$, namely $E^{\infty}\! =\! \argmin_{E\in[0,v]} i_{\rm RS}(E;\Delta)$. 
\end{corollary}
$\Delta_{\rm AMP}$ can be read off the replica potential
\eqref{eq:potentialfunction}: by differentiation of
\eqref{eq:potentialfunction} one finds a fixed point equation
that corresponds to \eqref{recursion-uncoupled-SE}. Thus $\Delta_{\rm AMP}$ is the smallest solution of 
$\partial i_{\rm RS}/ \partial E\!=\! \partial^2 i_{\rm RS}/\partial E^2 \!=\! 0$; in other words it is the ``first'' horizontal inflexion point that appears in $i_{\rm RS}(E;\Delta)$ when we increase $\Delta$.
\subsection{Discussion}
 
With our hypothesis on $P_0$ there are only three possible scenarios: 
$\Delta_{\rm AMP}\! <\!\Delta_{\rm RS}$ (one ``first order'' phase transition); $\Delta_{\rm AMP}\!=\!\Delta_{\rm RS}\! <\!+\infty$ (one ``higher order'' phase transition);
$\Delta_{\rm AMP}\!=\!\Delta_{\rm RS}\!=\! +\infty$ (no phase transition). 
In the sequel we will have in mind the most interesting case, namely {\it one first order phase transition}, where we determine 
the gap between the algorithmic AMP and information theoretic performance.
The cases of no phase transition
or higher order phase transition, which present no algorithmic gap, are basically covered by the analysis
of [\cite{6875223}] and follow as a special case from
our proof. The only cases that would require more work are those where $P_0$ is such that 
\eqref{eq:potentialfunction} develops more than three stationary points and more than 
one phase transition is present.  

\begin{figure}[!t]
\centering
\includegraphics[width=1\textwidth]{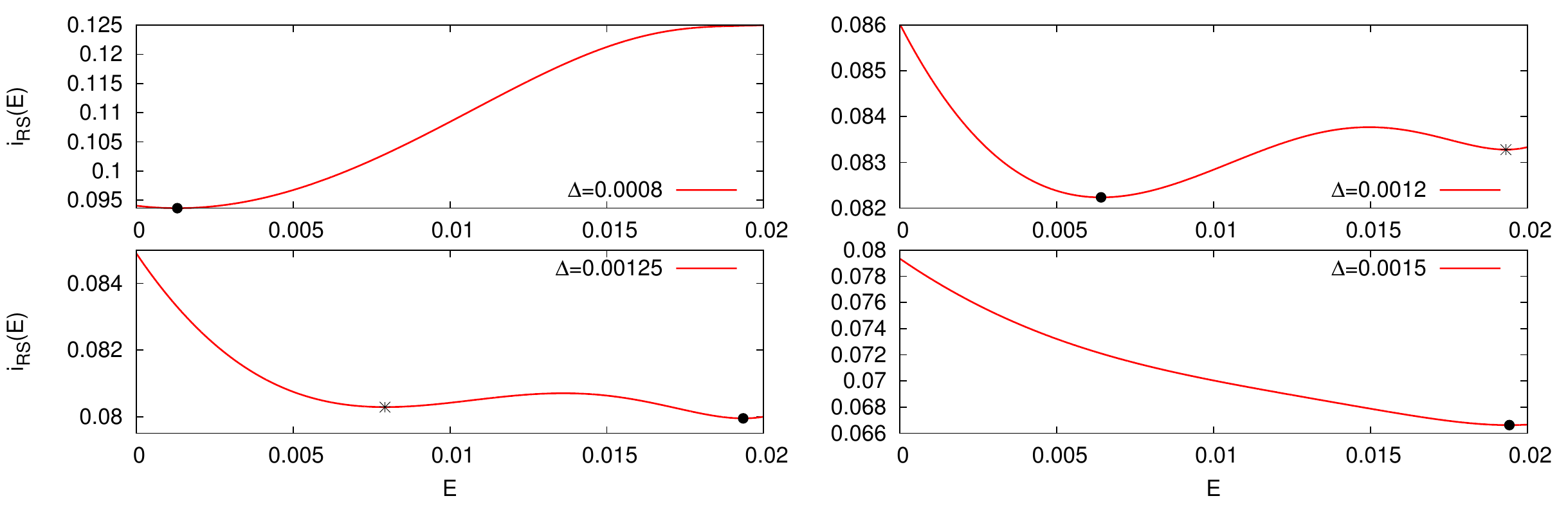}
\vspace*{-5pt}
\caption{\footnotesize The replica formula $i_{\rm RS}(E)$ for four values of $\Delta$ in the Wigner spike
  model. The mutual information is $\min i_{\rm RS}(E)$ (the black dot, while the black cross corresponds to the local minimum) and the asymptotic matrix-MMSE is $v^2\!-\!(v\!-\!\argmin_E i_{\rm RS}(E))^2$, where $v\!=\!\rho$ in this case with $\rho\!=\!0.02$ as in the inset of figure~\ref{fig-phase}. From
  top left to bottom right: \tbf{(1)} For low noise values, here
  $\Delta\!=\!0.0008\!<\!\Delta_{\rm AMP}$, there exists a unique ``good'' minimum corresponding to the MMSE and AMP is Bayes optimal. \tbf{(2)} As the noise
  increases, a second local ``bad'' minimum appears: this is the situation at
  $\Delta_{\rm AMP}\!<\!\Delta\!=\!0.0012\!<\!\Delta_{\rm RS}$. \tbf{(3)} For
  $\Delta\!=\!0.00125\!>\!\Delta_{\rm RS}$, the ``bad'' minimum
  becomes the global one and the MMSE suddenly deteriorates. \tbf{(4)} For
  even larger values of $\Delta$, only the
  ``bad'' minimum exists. The AMP algorithm
  can be seen as a naive minimizer of this curve starting from
  $E\!=\!v\!=\!0.02$. It reaches the global minimum in situations
  (1), (3) and (4), but in (2), when
  $\Delta_{\rm AMP}\!<\!\Delta\!<\!\Delta_{\rm RS}$, it is trapped by the
  local minimum with large MSE instead of reaching the global one corresponding to the MMSE.}\label{fig:I-AMP}
\end{figure}

For $\Delta_{\rm AMP}\!<\!\Delta_{\rm RS}$ the structure of stationary points of \eqref{eq:potentialfunction} is as follows\footnote{We take 
$\mathbb{E}[S]\!\neq \!0$. Once theorem~\ref{thm1} is proven for this case a limiting argument allows to extend it to $\mathbb{E}[S]\!=\!0$.}
(figure~\ref{fig:I-AMP}).
There exist three branches $E_{\rm good}(\Delta)$, 
$E_{\rm unstable}(\Delta)$ and $E_{\rm bad}(\Delta)$ such that:
{\bf 1)} For $0\!<\!\Delta\!<\!\Delta_{\rm AMP}$ there is a single stationary point $E_{\rm good}(\Delta)$ which is a global minimum;
{\bf 2)} At $\Delta_{\rm AMP}$ a {\it horizontal inflexion point} appears, for  
$\Delta\!\in\! [\Delta_{\rm AMP}, \Delta_{\rm RS}]$ there are three stationary points  
satisfying $E_{\rm good} (\Delta_{\rm AMP})\! <\! E_{\rm unstable}(\Delta_{\rm AMP})\! =\! E_{\rm bad}(\Delta_{\rm AMP})$, 
$E_{\rm good}(\Delta)\! < \!E_{\rm unstable}(\Delta)\! <\! E_{\rm bad}(\Delta)$ otherwise, and 
moreover $i_{\rm RS}(E_{\rm good}; \Delta)\! \leq\! i_{\rm RS}(E_{\rm bad}; \Delta)$ 
{\it with equality only at} $\Delta_{\rm RS}$;
{\bf 3)} for $\Delta \!>\! \Delta_{\rm RS}$ there is {\it at least} the stationary point $E_{\rm bad}(\Delta)$
which is always the global minimum, i.e. $i_{\rm RS}(E_{\rm bad}; \Delta) \!<\! i_{\rm RS}(E_{\rm good}; \Delta)$.
(For higher $\Delta$ the $E_{\rm good}(\Delta)$ and $E_{\rm unstable}(\Delta)$ branches may
merge and disappear); {\bf 4)} $E_{\rm good}(\Delta)$ is analytic for $\Delta\!\in ]0, \Delta^\prime[$, $\Delta^\prime \!>\!\Delta_{\rm RS}$, and 
$E_{\rm bad}(\Delta)$ is analytic for $\Delta \!>\! \Delta_{\rm AMP}$.  

We note for further use in the proof section that $E^\infty\! =\! E_{\rm good}(\Delta)$ for $\Delta\! <\! \Delta_{\rm AMP}$ 
and $E^\infty \!=\! E_{\rm bad}(\Delta)$ for $\Delta \!>\! \Delta_{\rm AMP}$. Definition \ref{algo-thresh-def} is equivalent to
$\Delta_{\rm AMP}\!=\!\sup\{\Delta\vert E^{\infty} \!=\! E_{\rm good}(\Delta)\}$. Moreover we will also use that 
$i_{\rm RS}(E_{\rm good}; \Delta)$ is analytic on $]0, \Delta^\prime[$, $i_{\rm RS}(E_{\rm bad}; \Delta)$ is analytic 
on $]\Delta_{\rm AMP}, +\infty[$, and the only non-analyticity point of $\min_{E\in [0,v]}i_{\rm RS}(E; \Delta)$ is at $\Delta_{\rm RS}$. 
\subsection{Relation to other works}
Explicit single-letter characterization of the mutual information in
the rank-one problem has attracted a lot of attention
recently. Particular cases of \eqref{eq:replicaformula} have
been shown rigorously in a number of situations. A special case when
$s_i\!=\!\pm 1\!\sim\! {\rm Ber}(1/2)$ already appeared in
[\cite{KoradaMacris}] where an equivalent spin glass model is analysed.
Very recently, [\cite{krzakala2016mutual}] has generalized the results
of [\cite{KoradaMacris}] and, notably, obtained a generic matching upper
bound. The same formula has been also rigorously computed following
the study of AMP in [\cite{6875223}] for spike models (provided,
however, that the signal was not {\it too} sparse) and in
[\cite{deshpande2015asymptotic}] for strictly symmetric community
detection. 

For rank-one symmetric matrix estimation problems, AMP has been
introduced by [\cite{rangan2012iterative}], who also computed the
state evolution formula to analyse its performance, generalizing
techniques developed by [\cite{BayatiMontanari10}] and
[\cite{Montanari-Javanmard}]. State evolution was further studied by
[\cite{6875223}] and [\cite{deshpande2015asymptotic}]. In
[\cite{lesieur2015phase,lesieur2015mmse}], the generalization to
larger rank was also considered.

The general formula proposed by [\cite{lesieur2015mmse}] for the
conditional entropy and the MMSE on the basis of the heuristic cavity
method from statistical physics was not demonstrated in full
generality. Worst, all existing proofs could not reach the more
interesting regime where a gap between the algorithmic and information
theoretic perfomances appears, leaving a gap with the statistical
physics conjectured formula (and rigorous upper bound from
[\cite{krzakala2016mutual}]). Our result closes this conjecture and
has interesting non-trivial implications on the computational
complexity of these tasks.

Our proof technique combines recent rigorous results in coding theory
along the study of capacity-achieving spatially coupled
codes [\cite{hassani2010coupled,kudekar2011threshold,6887298,DBLP:journals/corr/BarbierDM16a}]
with other progress, coming from developments in mathematical
physics putting on a rigorous basis predictions of spin glass
theory [\cite{guerra2005introduction}]. From this point of view, the
theorem proved in this paper is relevant in a broader context going
beyond low-rank matrix estimation. Hundreds of papers have been
published in statistics, machine learning or information theory using
the non-rigorous statistical physics approach. We believe that our
result helps setting a rigorous foundation of a broad line of
work. While we focus on rank-one symmetric matrix estimation, our
proof technique is readily extendable to more generic low-rank
symmetric matrix or low-rank symmetric tensor estimation. We also
believe that it can be extended to other problems of interest in
machine learning and signal processing, such as generalized linear
regression, features/dictionary learning, compressed sensing or
multi-layer neural networks.
\section{Two examples: Wigner spike model and community detection}
%
%
\begin{figure}[!t]
\centering
\includegraphics[width=0.50\textwidth]{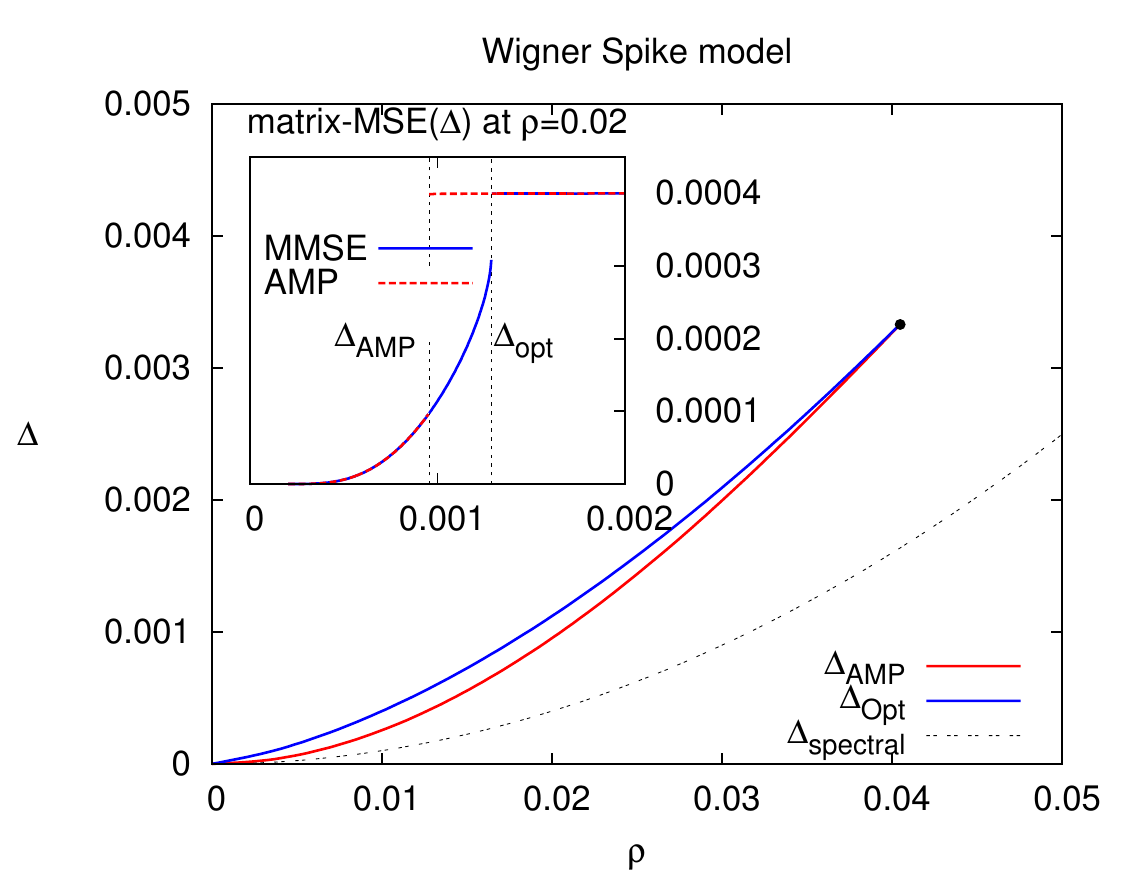}
\hspace{-10pt}
\includegraphics[width=0.50\textwidth]{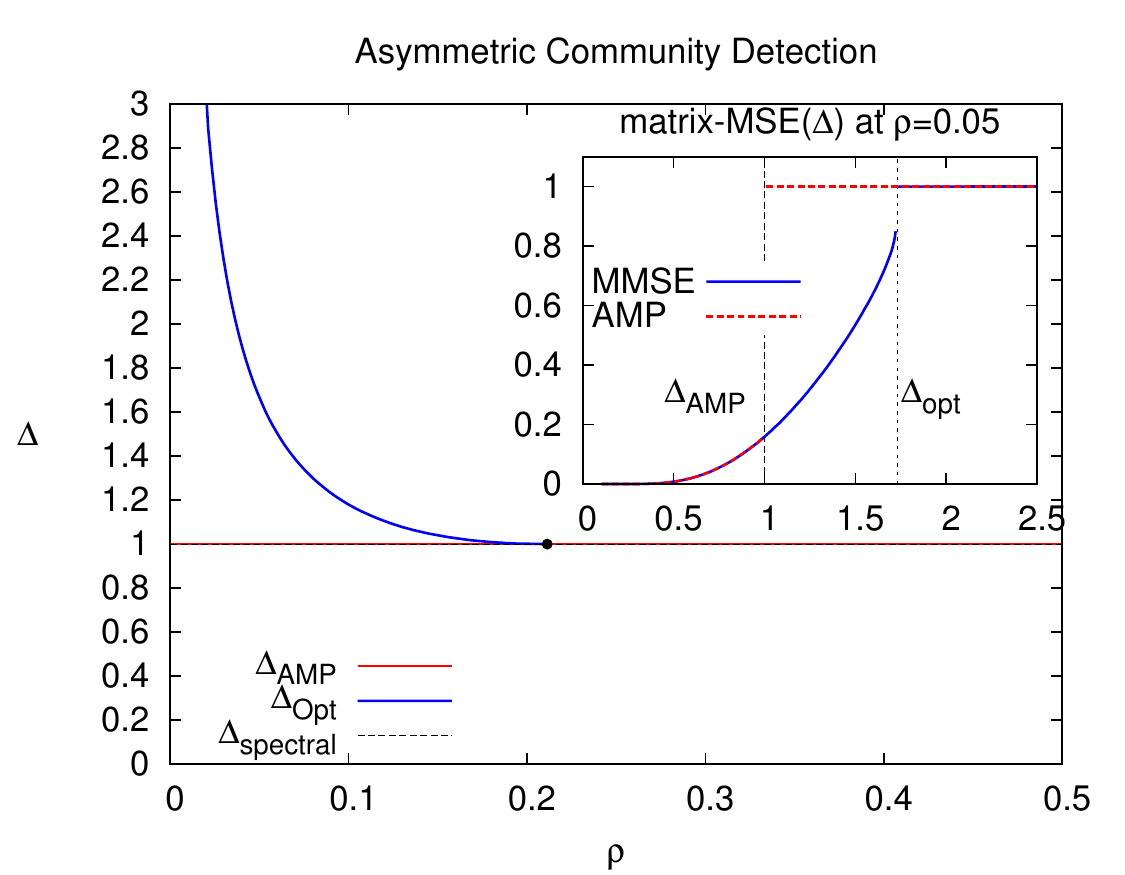}
\vspace*{-5pt}
\caption{\footnotesize Phase diagram in the noise variance $\Delta$ versus density $\rho$ plane for the rank-one spiked Wigner model (left) and
  the asymmetric community detection (right). \tbf{Left:} [\cite{6875223}]
  proved that AMP achieves the matrix-MMSE for all $\Delta$ as long as $\rho\!>\!0.041(1)$. Here we show that AMP is
  actually achieving the optimal reconstruction in the whole phase
  diagram except in the small region between the blue and red
  lines. Notice the large gap with spectral methods (dashed black
  line). \tbf{Inset:} matrix-MMSE (blue) at $\rho\!=\!0.02$ as a function of $\Delta$. AMP (dashed red) provably achieves the matrix-MMSE
  except in the region $\Delta_{\rm AMP}\!<\!\Delta\!<\!\Delta_{\rm Opt}\!=\!\Delta_{\rm RS}$. We conjecture that no polynomial-time algorithm will do better than
  AMP in this region. \tbf{Right:} Asymmetric community detection problem
  with two communities. For $\rho\!>\!1/2\!-\!\sqrt{1/12}$ (black point) and when $\Delta\!>\!1$, it is information theoretically impossible to
  find any overlap with the true communities and the matrix-MMSE is $1$, while it becomes possible for $\Delta\!<\!1$. In this region, AMP
  is always achieving the matrix-MMSE and spectral methods can find a
  non-trivial overlap with the truth as well, starting from
  $\Delta\!<\!1$. For $\rho\!<\!1/2\!-\!\sqrt{1/12}$, however, it is
  information theoretically possible to find an overlap with the hidden
  communities for $\Delta\!>\!1$ (below the blue line) but both AMP
  and spectral methods miss this information. \tbf{Inset:} matrix-MMSE (blue)
  at $\rho\!=\!0.05$ as a function of $\Delta$. AMP (dashed red) again provably achieves the matrix-MMSE except in the region
  $\Delta_{\rm AMP}\!<\!\Delta\!<\!\Delta_{\rm Opt}$.}
\label{fig-phase}
\end{figure}
In order to illustrate the consequences of our results we shall
present two examples. In the first one we are given data distributed
according to the spiked Wigner model where the vector $\tbf{s}$ is a Bernoulli random vector, $S_i\! \sim\!{\rm Ber}(\rho)$. For 
large enough densities (i.e. $\rho\!>\!0.041(1)$), [\cite{6875223}] computed the matrix-MMSE and 
proved that AMP is a computationally efficient algorithm that asymptotically achieves the matrix-MMSE 
for {\it any} value of the noise $\Delta$. Our results allow to close the gap left open by
[\cite{6875223}]: on one hand we now obtain rigorously the MMSE for
$\rho\!\le\!0.041(1)$, and on the other one, we observe that for such
values of $\rho$, and as $\Delta$ decreases, there is a small region
where two local minima coexist in $i_{\rm RS}(E; \Delta)$. In
particular for $\Delta_{\rm AMP}\!<\!\Delta\!<\!\Delta_{\rm Opt}=\Delta_{\rm RS}$ the
global minimum corresponding to the MMSE differs from the local one
that traps AMP, and a computational gap appears (see figure~\ref{fig:I-AMP}). While the region where
AMP is Bayes optimal is quite large, the region where is it not,
however, is perhaps the most interesting one. While this is by no
means evident, statistical physics analogies with physical phase
transitions in nature suggest that this region should be hard for a
very broad class of algorithms. 

For small $\rho$ our results are consistent with the known optimal and
algorithmic thresholds predicted in sparse PCA
[\cite{amini2008high,berthet2013computational}], that treats the case of
sub-extensive $\rho \!=\!\smallO(1)$ values. Another interesting line of
work for such probabilistic models appeared in the context of random
matrix theory (see [\cite{baik2005phase}] and references therein) and
predicts that a sharp phase transition occurs at a critical value of
the noise $\Delta_{\rm spectral} \!=\!\rho^2$ below which an outlier
eigenvalue (and its principal eigenvector) has a positive correlation
with the hidden signal. For larger noise values the spectral
distribution of the observation is indistinguishable from that of the
pure random noise.
%
%

We now consider the problem of detecting two communities (groups) with
different sizes $\rho n$ and $(1\!-\!\rho) n$, that generalizes the one
considered in [\cite{deshpande2015asymptotic}]. One is given a graph
where the probability to have a link between nodes in the first group
is $p\!+\! \mu (1\!-\!\rho)/(\rho\sqrt{n})$, between those in the second
group is $p\!+\!\mu\rho/(\sqrt{n}(1\!-\!\rho))$, while interconnections appear
with probability $p\!-\!\mu/\sqrt{n}$. With this peculiar ``balanced''
setting, the nodes in each group have the same degree distribution with
mean $pn$, making them harder to distinguish. According to the universality property described in section~\ref{sec:universal}, this is equivalent to a model with AWGN of variance $\Delta\!=\!p(1\!-\!p)/\mu^2$ where each variable $s_i$ is chosen according to
$
P_0(s)\!=\!\rho \delta(s\!-\!\sqrt{(1\!-\!\rho)/\rho}) \!+\!
(1\!-\!\rho)\delta(s\!+\!\sqrt{\rho/(1\!-\!\rho)}).
$
Our results for this problem\footnote{Note that here since $E\!=\!v\!=\!1$ is an extremum of $i_{\rm RS}(E;\Delta)$,
one must introduce a small bias in $P_0$ and let it then tend to zero at the end of the proofs.} 
are
summarized on the right hand side of figure~\ref{fig-phase}.
For
$\rho\!>\rho_c\!=\!1/2\!-\!\sqrt{1/12}$ (black point), it is
asymptotically information theoretically {\it possible} to get an
estimation better than chance if and only if $\Delta\!<\!1$. When
$\rho\!<\!\rho_c$, however, it becomes possible for much larger values
of the noise. Interestingly, AMP and spectral methods have the same
transition and can find a positive correlation with the hidden
communities for $\Delta\!<\!1$, regardless of the value of $\rho$. Again,
a region $[\Delta_{\rm AMP},\Delta_{\rm Opt}\!=\!\Delta_{\rm RS}]$ exists where a computational gap appears when $\rho\!<\!\rho_c$.

One can investigate the very low $\rho$ regime where we find that the
information theoretic transition goes as $\Delta_{\rm Opt}(\rho \!\to\!0)\!=\! 1/(4\rho\vert\log{\rho}\vert)$. Now if we assume that this result stays
true even for $\rho\!=\!\smallO(1)$ (which is a speculation at this point), we can choose $\mu\!\to\! (1\!-\!p)\rho\sqrt{n}$ such
that the small group is a clique. Then the problem corresponds to a ``balanced'' version of the famous planted clique problem
[\cite{d2007direct}]. We find that the AMP/spectral approach finds the hidden clique when it is larger than 
$\sqrt{np/(1\!-\!p)}$, while the information theoretic transition
translates into size of the clique 
$4p \log(n)/(1\!-\!p)$. This is indeed reminiscent of the more classical planted
clique problem at $p\!=\!1/2$ with its gap between $\log(n)$ (information
theoretic), $\sqrt{n/e}$ (AMP [\cite{deshpande2015finding}]) and
$\sqrt{n}$ (spectral [\cite{d2007direct}]). Since in our balanced case
the spectral and AMP limits match, this suggests that the small
gain of AMP in the standard clique problem is simply due to the
information provided by the distribution of local degrees in the two
groups (which is absent in our balanced case). We believe this correspondence strengthens the claim that the AMP gap is actually a fundamental one.
\section{Proofs}
The crux of our proof rests on an auxiliary ``spatially coupled system''. The hallmark of spatially coupled models is that one can tune them so that 
the gap between the algorithmic and information theoretical limits can be eliminated, while at the same time the mutual information is maintained unchanged for the coupled and original models. Roughly speaking, this means that it is possible to algorithmically compute 
the information theoretical limit of the original model because a
suitable algorithm is optimal on the coupled system. 

The spatially coupled construction used here is very similar to the
one used for the coupled Curie-Weiss model [\cite{hassani2010coupled}].
We consider a ring of length $L\!+\!1$ ($L$ even) with {\it blocks} positioned at
$\mu\!\in\! \{0,\dots, L\}$ and coupled to neighboring blocks
$\{\mu\! -\! w, \dots, \mu\! +\!w\}$. The positions $\mu$ are taken modulo $L\!+\!1$ and $w\!\in\!\{0,\ldots,L/2\}$ is an integer equal to the size of the {\it coupling window}. The coupled model is
\begin{align} \label{eq:def_coupledSyst}
w_{i_\mu j_\nu} = s_{i_\mu} s_{j_\nu}\sqrt{\frac{\Lambda_{\mu\nu}}{n}} + z_{i_\mu j_\nu}\sqrt{\Delta} ,
\end{align}
where the index $i_\mu \!\in\! \{1, \dots, n\}$ (resp. $j_\nu$) belongs to the block $\mu$ (resp. $\nu$) along the ring, $\mathbf{\Lambda}$ is an $(L\!+\!1)\!\times\! (L\!+\!1)$ matrix which describes the strength of the coupling between blocks, and $Z_{i_\mu j_\nu}\!\sim \!\mathcal{N}(0,1)$ are i.i.d. For the proof to work, the matrix elements have to be chosen appropriately. We assume that: $i)$ $\mathbf{\Lambda}$ is a doubly stochastic matrix; $ii)$ $\Lambda_{\mu\nu}$ depends on
$\vert \mu \!- \!\nu\vert$; $iii)$ $\Lambda_{\mu\nu}$ is not vanishing for $\vert \mu \!-\!\nu| \leq w$ and vanishes for $\vert \mu \!-\!\nu|\! > \!w$; $iv)$ $\mathbf{\Lambda}$ is {\it smooth} in the sense $\vert \Lambda_{\mu\nu}\! -\! \Lambda_{\mu+1 \nu} \vert\! =\! \mathcal{O}(w^{-2})$; $v)$ $\mathbf{\Lambda}$ has a non-negative Fourier transform. All these conditions can easily be met, the simplest example being a triangle of base $2w\!+\!1$ and height $1/(w\!+\!1)$. 
The construction of the coupled system is completed by introducing a {\it seed} in the ring: we assume perfect knowledge of the signal components $\{s_{i_\mu}\}$ for $\mu\! \in \!\mathcal{B}\!\defeq \!\{-w\!-\!1, \dots, w\!-\!1\} \mod L\!+\!1$. This seed is what allows to close the 
gap between the algorithmic and information theoretical limits and therefore plays a crucial role. Note it can also be viewed as an ``opening'' 
of the chain with pinned boundary conditions. 

Our first crucial result states that the mutual information $I_{w, L}(\tbf{S}; \tbf{W})$ of the coupled and original systems are the same
in a suitable asymptotic limit.
%
\begin{lemma}[Equality of mutual informations]\label{LemmaGuerraSubadditivityStyle}
 For any $w\!\in\!\{0,\ldots,L/2\}$ the following limits exist and are equal: $\lim_{L\to +\infty}\lim_{n\to +\infty} I_{w, L}(\tbf{S}; \tbf{W})/(n(L\!+\!1))\! = \!\lim_{n\to +\infty} I(\tbf{S}; \tbf{W})/n$.
\end{lemma}
An immediate corollary is that non-analyticity points (w.r.t $\Delta$) of the mutual informations 
are the same in the coupled and original models.
In particular, defining $\Delta_{{\rm Opt, coup}}\! \defeq \!
\sup\{\Delta\mid \lim_{L\to +\infty}\lim_{n\to +\infty} I_{w, L}(\tbf{S}; \tbf{W})/(n(L\!+\!1)) \
\text{is analytic in}\ ]0, \Delta[\}$, we have 
$\Delta_{{\rm Opt, coup}} \!= \!\Delta_{{\rm Opt}}$.

The second crucial result states that the AMP threshold of the spatially coupled system is at least as good as $\Delta_{\rm RS}$. The analysis of
AMP applies to the coupled system as well
[\cite{BayatiMontanari10,Montanari-Javanmard}] and it can be shown that
the performance of AMP is assessed by state evolution. Let $E_{\mu}^{t} \!\defeq\!\lim_{n\to+\infty}\mathbb{E}_{{\tbf S}, {\tbf Z}}[\|{\tbf S}_{\mu} \!-\! \hat{\bs}_{\mu}^t\|_2^2]/n$ be the asymptotic average vector-MSE of the AMP estimate
$\hat{\tbf{\bs}}_\mu^t$ at time $t$ for the $\mu$-th ``block'' of ${\tbf S}$. We associate to each position $\mu\!\in\! \{0, \ldots, L\}$ an \emph{independent} scalar system with AWGN noise of the form $Y\!=\!S\!+\!\Sigma_\mu({\tbf E}; \Delta)Z$ with $\Sigma_\mu({\tbf E}; \Delta)^2 \!\defeq\! \Delta/(v\!-\! \sum_{\nu=0}^L\Lambda_{\mu\nu} E_\nu)$ and $S\!\sim\! P_0$, $Z\!\sim \!\mathcal{N}(0,1)$. Taking into account knowledge of the signal in $\mathcal{B}$, state evolution reads:
%
\begin{align}\label{coupled-state-evolution}
 E_{\mu}^{t+1} = {\rm mmse}(\Sigma_\mu({\tbf E}^t; \Delta)^{-2}), ~ E_\mu^0 = v ~\text{for}~ \mu\in \{0,\ldots, L\}\setminus\mathcal{B}, ~
 E_\mu^t = 0 ~\text{for}~ \mu\in \mathcal{B}, t\geq 0,
\end{align}
where the mmse function is defined as in section~\ref{subsec:mainresult}. From the monotonicity of the mmse function we have 
 $E_\mu^{t+1} \!\leq\! E_\mu^t$ for all $\mu\!\in\! \{0, \ldots, L\}$, a partial order which implies that $\lim_{t\to+\infty} {\tbf E}^t \!=\! {\tbf E}^\infty$ exists. This allows to define an algorithmic threshold:
$\Delta_{{\rm AMP}, w,L}\!\defeq\! \sup\{\Delta|E^{\infty}_\mu\!\le\! E_{\rm good}(\Delta) \ \forall \ \mu\}$. We show (equality holds but is not directly needed)
\begin{lemma}[Threshold saturation]\label{thres-sat-lemma} 
Let 
$\Delta_{\rm AMP, coup} \!\defeq \!\liminf_{w\to +\infty}\liminf_{L\to +\infty} \Delta_{{\rm AMP}, w, L}$.
We have $\Delta_{\rm AMP, coup} \!\geq \!\Delta_{\rm RS}$. 
\end{lemma}
%
%

\noindent{\bf Proof sketch of theorem~\ref{thm1}} {\it First we prove \eqref{eq:replicaformula} for} $\Delta \leq \Delta_{\rm Opt}$.
It is known [\cite{6875223}] that the matrix-MSE of AMP when $n\!\to \!+\infty$ is equal to $v^2 \!-\!(v \!-\! E^t)^2$. This cannot improve the matrix-MMSE, hence
\begin{align}\label{ineq}
 \frac{1}{4}(v^2 -(v - E^\infty)^2) \geq \limsup_{n\to +\infty}\frac{1}{4n^2}\mathbb{E}_{\tbf{S}, \tbf{W}}\|\tbf{S}\tbf{S}^{\intercal} - \mathbb{E}[\tbf{X}\tbf{X}^{\intercal}\vert \tbf{W}]\|_{\rm F}^2.
\end{align}
For $\Delta \!\leq\! \Delta_{\rm AMP}$ we have $E^{\infty}\!=\! E_{\rm good}(\Delta)$ which is
 the global minimum of \eqref{eq:potentialfunction} so the left hand side of \eqref{ineq} is equal to the derivative of 
 $\min_{E\in [0,v]}i_{\rm RS}(E;\Delta)$ w.r.t $\Delta^{-1}$. Thus using a matrix version of the well known I-MMSE relation [\cite{GuoShamaiVerdu}]
we get 
\begin{align}\label{important}
 \frac{d}{d\Delta^{-1}}\min_{E\in [0,v]} i_{\rm RS}(E;\Delta) \geq \limsup_{n\to +\infty}\frac{1}{n} \frac{d I(\tbf{S}; \tbf{W})}{d\Delta^{-1}}.
\end{align}
Integrating this relation on $[0, \Delta] \!\subset\![0, \Delta_{\rm AMP}]$ and checking that $\min_{E\in [0,v]}i_{\rm RS}(E; 0)\! =\! H(S)$ (the Shannon entropy
of $P_0$) we obtain
$\min_{E\in [0,v]} i_{\rm RS}(E;\Delta) \!\leq\! \liminf_{n\to +\infty}I(\tbf{S}; \tbf{W})/n$. But we know 
$I(\tbf{S}; \tbf{W})/n \!\leq\! \min_{E\in [0,v]} i_{\rm RS}(E; \Delta)$ [\cite{krzakala2016mutual}], thus we already get
 \eqref{eq:replicaformula} for $\Delta\! \leq\! \Delta_{\rm AMP}$. We notice that $\Delta_{\rm AMP} \!\leq\! \Delta_{\rm Opt}$. While this might seem intuitively clear, it follows from $\Delta_{\rm RS} \!\geq \!\Delta_{\rm AMP}$ (by their definitions) which together with 
 $\Delta_{\rm AMP} \!>\! \Delta_{\rm Opt}$ would imply from \eqref{eq:replicaformula} that $\lim_{n\to +\infty} I(\tbf{S}; \tbf{W})/n$ is analytic at $\Delta_{\rm Opt}$, a contradiction. The next step is 
 to extend \eqref{eq:replicaformula} to the range $[\Delta_{\rm AMP}, \Delta_{\rm Opt}]$. 
Suppose for a moment $\Delta_{\rm RS}\! \geq \!\Delta_{\rm Opt}$. Then 
{\it both} functions on each side of \eqref{eq:replicaformula} are analytic on the whole range $]0, \Delta_{\rm Opt}[$ and since they are equal for $\Delta \!\leq \!\Delta_{\rm AMP}$, they {\it must} be equal on their whole analyticity range and by continuity, they must also be equal at $\Delta_{\rm Opt}$ (that the functions are continuous follows from independent arguments on the existence of the $n\!\to\!+\infty$ limit of concave functions). It remains to show that $\Delta_{\rm RS}\!\in \,]\Delta_{\rm AMP},  \Delta_{\rm Opt}[$ is impossible. We proceed by contradiction, so suppose this is true.
Then both functions on each side of \eqref{eq:replicaformula} are analytic on $]0, \Delta_{\rm RS}[$ and since they are equal for $]0, \Delta_{\rm AMP}[ \subset ]0, \Delta_{\rm RS}[$ they must 
be equal on the whole range $]0, \Delta_{\rm RS}[$ and also at $\Delta_{\rm RS}$ by continuity. For $\Delta\!> \!\Delta_{\rm RS}$ the fixed point of state evolution is $E^{\infty} \!= \!E_{\rm bad}(\Delta)$ which is also the global minimum of $i_{\rm RS}(E;\Delta)$, hence 
\eqref{important} is verified. Integrating this inequality on $]\Delta_{\rm RS}, \Delta[ \subset ]\Delta_{\rm RS}, \Delta_{\rm Opt}[$ and using
$I(\tbf{S}; \tbf{W})/n \!\leq\! \min_{E\in [0,v]}i_{\rm RS}(E; \Delta)$ again, we find that 
\eqref{eq:replicaformula} holds for all $\Delta\!\in\! [0, \Delta_{\rm Opt}]$. But this implies that $\min_{E\in [0,v]} i_{\rm RS}(E;\Delta)$ is analytic at $\Delta_{\rm RS}$, a contradiction.

{\it We now prove \eqref{eq:replicaformula} for} $\Delta \!\geq \!\Delta_{\rm Opt}$. Note that the previous arguments showed that necessarily
$\Delta_{\rm Opt} \!\leq \!\Delta_{\rm RS}$.
Thus by lemmas~\ref{LemmaGuerraSubadditivityStyle} and \ref{thres-sat-lemma} (and the sub-optimality of AMP as shown as before) we obtain $\Delta_{\rm RS}\leq \Delta_{\rm AMP, coup}\!\leq\! \Delta_{\rm Opt, coup}\!
=\! \Delta_{\rm Opt} \!\leq\! \Delta_{\rm RS}$. 
This shows that $\Delta_{\rm Opt} \!=\! \Delta_{\rm RS}$ (this is the point where spatial coupling came in the game and 
 we do not know of other means to prove such an equality). For $\Delta \!>\! \Delta_{\rm RS}$ we have $E^\infty \!=\! E_{\rm bad}(\Delta)$ which is the global minimum of $i_{\rm RS}(E;\Delta)$. Therefore we again have \eqref{important} in this range and the proof can be completed 
by using once more the integration argument, this time over the range $[\Delta_{\rm RS}, \Delta] \!=\! [\Delta_{\rm Opt}, \Delta]$.

\noindent{\bf Proof sketch of corollaries~\ref{cor:MMSE} and \ref{perfamp}} 
Let $E_*(\Delta)\! =\! \argmin_E i_{\rm RS}(E;\Delta)$ for $\Delta\!\neq\! \Delta_{\rm RS}$. 
By explicit calculation one checks that $d i_{\rm RS}(E_*, \Delta)/d\Delta^{-1}\!=\!(v^2\! - \!(v\!-\! E_*(\Delta))^2)/4$, so from
theorem~\ref{thm1} and the matrix form of the I-MMSE relation we find 
${\rm Mmmse}_n \!\to\! v^2\! -\! (v\!-\! E_*(\Delta))^2$
as $n\!\to \!+\infty$ which is the first part of the statement of corollary~\ref{cor:MMSE}.
Let us now turn to corollary~\ref{perfamp}.
For $n\!\to\! +\infty$ the vector-MSE of the AMP estimator at time $t$ equals 
$E^t$, and since the fixed point equation corresponding to state evolution 
is precisely the stationarity equation for $i_{\rm RS}(E;\Delta)$, we conclude that 
for $\Delta\!\notin\! [\Delta_{\rm AMP}, \Delta_{\rm RS}]$ we must 
have $E^\infty\! = \!E_*(\Delta)$. It remains to prove that $E_*(\Delta) \!= \!\lim_{n\to +\infty}{\rm Vmmse}_n(\Delta)$
at least for $\Delta\!\notin \![\Delta_{\rm AMP}, \Delta_{\rm RS}]$ (we believe this is in fact true for all $\Delta$). 
This will settle the second part of corollary~\ref{cor:MMSE} as well
as \ref{perfamp}.
Using
(Nishimori) identities $\mathbb{E}_{\tbf{S}, \tbf{W}}[S_iS_j \mathbb{E}[X_iX_j\vert \tbf{W}]]\!= 
\!\mathbb{E}_{\tbf{S}, \tbf{W}}[\mathbb{E}[X_iX_j\vert \tbf{W}]^2]$
(see e.g. [\cite{krzakala2016mutual}]) and the law of large numbers we can show
$\lim_{n\to +\infty}{\rm Mmmse}_n \!\leq\! \lim_{n\to +\infty} (v^2\! -\! (v\!-\!{\rm Vmmse}_n(\Delta))^2)$.
Concentration techniques similar to [\cite{KoradaMacris}] suggest that the equality in fact holds
(for $\Delta\!\neq \!\Delta_{\rm RS}$) but 
there are technicalities that prevent us 
from completing the proof of equality. However it is interesting to note that this equality 
would imply $E_*(\Delta) \! =\!\lim_{n\to +\infty}{\rm Vmmse}_n(\Delta)$ for all $\Delta\!\neq\! \Delta_{\rm RS}$.
Nevertheless, another argument can be used when AMP is optimal.
On one hand the right hand side of the inequality is necessarily smaller than $v^2 \!- \!(v\!-\! E^\infty)^2$. On the other hand 
the left hand side of the inequality is equal to $v^2\! -\! (v\!-\! E_*(\Delta))^2$. Since $E_*(\Delta)\!=\!E^\infty$ when 
$\Delta\!\notin\! [\Delta_{\rm AMP}, \Delta_{\rm RS}]$, we can conclude $\lim_{n\to +\infty}{\rm Vmmse}_n(\Delta) \!=\! \argmin_E i_{\rm RS}(E;\Delta)$ for this range of $\Delta$.  
\\

\noindent{\bf Proof sketch of lemma~\ref{LemmaGuerraSubadditivityStyle}} Here we prove the lemma for a ring that is not seeded. An easy argument
 shows that a seed of size $w$ does not change the mutual information per variable when $L\!\to\! +\infty$.
The statistical physics formulation is convenient: up to a trivial additive term equal to $n(L\!+\!1)v^2/4$, the mutual information $I_{w, L}(\tbf{S}; \tbf{W})$ is equal to the 
free energy $-\mathbb{E}_{\tbf{S}, \tbf{Z}}[\ln \mathcal{Z}_{w, L}]$, where 
$\mathcal{Z}_{w,L}\!\defeq\! \int d{\tbf x}P_0({\tbf x}) \exp(-\mathcal{H}({\tbf x},{\tbf z}, \mathbf{\Lambda}))$ is the partition function with Hamiltonian 
\begin{align}\label{interp-ham}
\mathcal{H}({\tbf x},{\tbf z}, \mathbf{\Lambda})&=\frac{1}{\Delta}
\sum_{\mu=0}^L \Lambda_{\mu\mu} \sum_{i_\mu\le j_\mu} \biggl(\frac{x_{i_\mu}^2x_{j_\mu}^2}{2n} - \frac{s_{i_\mu}s_{j_\mu}x_{i_\mu}x_{j_\mu}}{n} - \frac{x_{i_\mu}x_{j_\mu}z_{i_\mu j_\mu}\sqrt{\Delta}}{\sqrt{n\Lambda_{\mu\mu}}}\biggr) \nonumber\\
&+ \frac{1}{\Delta}\sum_{\mu=0}^L \sum_{\nu=\mu+1}^{\mu+w} \Lambda_{\mu\nu} \sum_{i_\mu, j_\nu} \biggl(\frac{x_{i_\mu}^2x_{j_\nu}^2}{2n} - \frac{s_{i_\mu}s_{j_\nu}x_{i_\mu}x_{j_\nu}}{n} - \frac{x_{i_\mu}x_{j_\nu}z_{i_\mu j_\nu}\sqrt{\Delta}}{\sqrt{n\Lambda_{\mu\nu}}}\biggr).
\end{align}
Consider a pair of systems with coupling matrices $\mathbf{\Lambda}$ and $\mathbf{\Lambda}^\prime$ and i.i.d noize realizations ${\tbf z},{\tbf z}^\prime$, an \emph{interpolated Hamiltonian} $\mathcal{H}({\tbf x},{\tbf z}, t\mathbf{\Lambda}) \!+ \!\mathcal{H}({\tbf x},{\tbf z}^\prime, (1\!-\!t)\mathbf{\Lambda}^\prime)$, $t\in [0,1]$, and the corresponding partition function $\mathcal{Z}_t$. The main idea of the proof is to show that for suitable choices of matrices, $-\frac{d}{dt}\mathbb{E}_{\tbf{S}, \tbf{Z},\tbf{Z}^\prime}[\ln \mathcal{Z}_{t}]$ is negative for all $t\!\in\! [0,1]$ (up to negligible terms), so that by the fundamental theorem of calculus, we get a comparison between the free energies of $\mathcal{H}({\tbf x},{\tbf z}, \mathbf{\Lambda})$ and
$\mathcal{H}({\tbf x},{\tbf z}^\prime ,\mathbf{\Lambda}^\prime)$. Performing the $t$-derivative brings down a Gibbs average of a polynomial in all
variables $s_{i_\mu}$, $x_{i_\mu}$, $z_{i_\mu j_\nu}$ and $z_{i_\mu j_\nu}^\prime$. This expectation over ${\tbf S}$, ${\tbf Z}$, ${\tbf Z}^\prime$ of this Gibbs average can be greatly simplified using integration by parts over the Gaussian noise $z_{i_\mu j_\nu}, z_{i_\mu j_\nu}^\prime$ and Nishimori identities (see e.g. proof of corollary~\ref{cor:MMSE} for one of them). This algebra leads to
\begin{align}
-\frac{1}{n(L+1)}\frac{d}{dt}\mathbb{E}_{\tbf{S}, \tbf{Z},{\tbf Z}^\prime}[\ln \mathcal{Z}_t] = 
\frac{1}{4\Delta (L+1)}\mathbb{E}_{\tbf{S}, \tbf{Z},{\tbf Z}^\prime}[\langle {\tbf q}^\intercal\mathbf{\Lambda} {\tbf q} - {\tbf q}^\intercal\mathbf{\Lambda}^\prime {\tbf q}\rangle_t]
+ \mathcal{O}(1/(nL)) ,
\end{align}
where $\langle \!-\!\rangle_t$ is the Gibbs average w.r.t the interpolated Hamiltonian, ${\tbf q}$ is the vector of overlaps $q_\mu \!\defeq\! \sum_{i_\mu=1}^n s_{i_\mu} x_{i_\mu}/n$. If we can choose matrices such that $\mathbf{\Lambda}^\prime \!> \!\mathbf{\Lambda}$, the difference of quadratic forms in the Gibbs bracket is negative and we obtain an inequality in the large size limit. We use this scheme to interpolate between the fully decoupled system $w\!=\!0$ and the coupled one $1\!\leq\! w\!<\! L/2$ and then between $1\!\leq \!w<\! L/2$ and the fully connected system $w\!=\!L/2$. 
The $w\!=\!0$ system has $\Lambda_{\mu\nu}\!=\!\delta_{\mu\nu}$ with eigenvalues $(1,1,\dots, 1)$. For the $1\!\leq \!w\!< \!L/2$ system, we take any 
stochastic translation invariant matrix with non-negative discrete Fourier transform (of its rows): such matrices have an eigenvalue
equal to $1$ and all others in $[0, 1[$ (the eigenvalues are precisely equal to the discrete Fourier transform). For $w\!=\!L/2$ we choose 
$\Lambda_{\mu\nu} \!=\! 1/(L\!+\!1)$ which is a projector with eigenvalues $(0, 0, \dots, 1)$. With these choices we deduce that the free energies and mutual informations are ordered as 
$I_{w=0, L} + \mathcal{O}(1)\! \leq \!I_{w, L} + \mathcal{O}(1) \!\leq\! I_{w=L/2, L} + \mathcal{O}(1)$. To conclude the proof we divide by $n(L\!+\!1)$ and note that the 
limits of the leftmost and rightmost mutual informations are equal, provided the limit exists. Indeed the leftmost term equals 
$L$ times $I(\tbf{S}; \tbf{W})$ and the rightmost term is the {\it same} mutual information for a system of $n(L\!+\!1)$ variables. 
Existence of the limit follows by a subadditivity inequality which itself is proven by a similar interpolation [\cite{guerra2005introduction}].  
\\

\noindent{\bf Proof sketch of lemma~\ref{thres-sat-lemma}} 
Fix $\Delta \!<\! \Delta_{\rm RS}$. We show  
that, for $w$ large enough, the coupled state evolution recursion \eqref{coupled-state-evolution} must converge to a fixed point
 $E_\mu^\infty \!\leq\! E_{\rm good}(\Delta)$ for all $\mu$. The main intuition behind the proof is to use a ``potential function'' whose ``energy'' can be lowered 
 by small perturbation of a fixed point that {\it would} go above $E_{\rm good}(\Delta)$ [\cite{6887298,DBLP:journals/corr/BarbierDM16a}]. The relevant potential function 
 $i_{w, L}({\tbf E}, \Delta)$ is in fact the replica potential of the coupled system, and equals up to a constant $(2w\!+\!1)Lv^2/4\Delta$ 
 \begin{align*}
\sum_{\mu}\biggl\{
\sum_{\nu=\mu -w}^{\mu+w}\frac{\Lambda_{\mu\nu}}{4\Delta}(v\!-\! E_\mu)(v\!-\!E_\nu)
\!-\!
\mathbb{E}_{{S}, {Z}}\biggl[ \ln \biggl(\int dx\, P_0(x) e^{-\frac{x^2}{2\Sigma_\mu({\tbf E};\Delta)^{2}} 
+ x\bigl(\frac{S}{\Sigma_\mu({\tbf E};\Delta)^{2}} + \frac{Z}{\Sigma_\mu({\tbf E};\Delta)}\bigr)} \biggr)  \biggr]
\biggr\}.
\end{align*}
We note that the stationarity condition for this potential is precisely \eqref{coupled-state-evolution} (without the seeding condition). 
Monotonicity properties of 
state evolution ensure that any fixed point has a ``unimodal'' shape (and recall that it vanishes for $\mu\!\in\! \mathcal{B}\!=\! \{0,\dots, w\!-\!1\} \cup\{L\!-\!w,\dots, L\}$). Consider a 
position $\mu_{\max}\! \in\! \{w,\dots, L\!-\!w\!-\!1\}$ where it is 
maximal and suppose that $E^{\infty}_{\mu_{\max}}\! >\! E_{\rm good}(\Delta)$.
We associate to the {\it fixed point} ${\tbf E}^{\infty}$ a so-called 
{\it saturated profile} ${\tbf E}^{\rm s}$ defined on the whole of $\mathbb{Z}$ as follows: $E_{\mu}^{\rm s} \!=\! E_{\rm good}(\Delta)$ for all $\mu\!\leq \!\mu_{\infty}$ where $\mu_{\infty}\!+\!1$ is the smallest position such that $E^{\infty}_{\mu} \!> \!E_{\rm good}(\Delta)$; $E^{\rm s}_\mu \!=\! E^{\infty}_\mu$ for $\mu\!\in\! \{\mu_{\infty}\!+\!1, \dots, \mu_{\max}\!-\!1\}$; 
$E^{\rm s}_\mu \!=\! E_{\mu_{\max}}^{\infty}$ for all $\mu\!\geq \!\mu_{\max}$. We show that ${\tbf E}^{\rm s}$ cannot exist for $w$ large enough.
To this end define a shift operator 
by $[\mathcal{S}({\tbf E}^{\rm s})]_\mu \!\defeq \!E^{\rm s}_{\mu-1}$. On one hand the shifted profile is a {\it small} perturbation of ${\tbf E}^{\rm s}$ which matches a fixed point, except 
where it is constant, so if we Taylor expand, the first order vanishes and the second order and higher orders can be estimated as 
$\vert i_{w, L}(\mathcal{S}({\tbf E}^{\rm s}); \Delta) \!- \!i_{w, L}({\tbf E}^{\rm s}; \Delta)\vert \!=\! \mathcal{O}(1/w)$ uniformly in $L$. 
On the other hand, by explicit cancellation of telescopic sums 
$i_{w, L}(\mathcal{S}({\tbf E}^{\rm s}); \Delta) \!-\! i_{w, L}({\tbf E}^{\rm s}; \Delta)\! 
=\! i_{\rm RS}(E_{\rm good};\Delta) \!-\! i_{\rm RS}(E_{\mu_{\max}}^{\infty};\Delta)$. Now one can show from monotonicity 
properties of state evolution that if ${\tbf E}^{\infty}$ is a fixed point then $E_{\mu_{\max}}^{\infty}$ \emph{cannot be in the basin of attraction of} $E_{\rm good}(\Delta)$ for the uncoupled recursion. Consequently as can be seen on the plot of $i_{\rm RS}(E;\Delta)$ (e.g. figure~\ref{fig:I-AMP}) we must have 
$i_{\rm RS}(E_{\mu_{\max}}^{\infty};\Delta)\!\geq \!i_{\rm RS}(E_{\rm bad};\Delta)$. 
Therefore $i_{w, L}(\mathcal{S}({\tbf E}^{\rm s}); \Delta) \!- \!i_{w, L}({\tbf E}^{\rm s}; \Delta)\!
\leq \!- \vert i_{\rm RS}(E_{\rm bad};\Delta)\! - \!i_{\rm RS}(E_{\rm good};\Delta)\vert$ which is an energy gain independent of $w$, and for large enough $w$ we get a contradiction with the previous estimate coming from the Taylor expansion.
\section*{Acknowledgments}
J.B and M.D acknowledge funding from the Swiss National Science
Foundation (grant num. 200021-156672).  Part of the research has
received funding from the European Research Council under the European
Union’s 7th Framework Programme (FP/2007-2013/ERC Grant Agreement
307087-SPARCS). This work was done in part while F.K and L.Z were
visiting the Simons Institute for the Theory of Computing.
\bibliographystyle{unsrt}
\bibliography{sample}
\end{document}

%% file: settings_custom.tex
\DeclareUnicodeCharacter{00A0}{~}

\def \({\left(}
\def \){\right)}
\def \[{\left[}
\def \]{\right]}
\newcommand{\argmin}{\text{argmin}}
\newcommand{\tbf}[1]{{\textbf{#1}}}

\newcommand{\defeq}{\vcentcolon=}

\newcommand{\bs}{{\textbf {s}}}

\newcommand{\be}{\begin{equation}}
\newcommand{\ee}{\end{equation}}
\newcommand{\bea}{\begin{eqnarray}}
\newcommand{\eea}{\end{eqnarray}}

\newcommand\smallO{
  \mathchoice
    {{\scriptstyle\mathcal{O}}}
    {{\scriptstyle\mathcal{O}}}
    {{\scriptscriptstyle\mathcal{O}}}
    {\scalebox{.7}{$\scriptscriptstyle\mathcal{O}$}}
  }